\documentclass[multphys,vecphys]{svmult}

\usepackage{makeidx}         
\usepackage{graphicx}        
\usepackage{multicol}        
\usepackage[bottom]{footmisc}

\makeindex             

\begin{document}

\title*{II Zwicky 23 and Family}
\author{Elizabeth H. Wehner\inst{1},
John S. Gallagher\inst{2},~Gwen C. Rudie\inst{3},\and Philip J. Cigan\inst{2}}
\authorrunning{Wehner et al.}
\institute{McMaster University, Department of Physics and Astronomy, Hamilton, ON  L8S 4M1, Canada
\texttt{wehnere@physics.mcmaster.ca} \and University of Wisconsin, 475 N. Charter St., Madison, WI 53706 \texttt{jsg@astro.wisc.edu, pjcigan@students.wisc.edu} \and Dartmouth College, Department of Physics and Astronomy, 6127 Wilder Laboratory, Hanover, NH, 03755 \texttt{gwen.c.rudie@dartmouth.edu}}
%
%
\maketitle

\begin{abstract}

II Zwicky 23 (UGC 3179) is a luminous ($M_B \sim -21$) nearby compact narrow emission line
st arburst galaxy with blue optical colors and strong emission lines. We present a
photometric and morphological study of II Zw 23 and its interacting companions
using data obtained with the WIYN 3.5-m telescope in Kitt Peak, Arizona.
II Zwicky 23 has a highly disturbed outer structure
with long trails of debris that may be feeding tidal dwarfs. Its central regions
appear disky, a structure that is consistent with the overall rotation pattern
observed in the H$\alpha$ velocity field measured from Densepak observations obtained
with WIYN. We discuss the structure of II Zwicky 23 and its set of companions and possible scenarios of debris formation in this system.

\end{abstract}

\section{Introduction}
\label{wehner:intro}

II Zwicky 23 (Mkn 1087, UGC 03179) is a blue compact galaxy currently undergoing a
global starburst.  It was until recently thought to have one companion, KPG103a. 
However, recent work\cite{wehner:ls04} has identified several tidal dwarfs around this
system, as well as one dwarf galaxy that appears to pre-date the tidal dwarf galaxies
(TDGs).  There are several indicators suggesting that these group members are
currently interacting.  The presence of tidally forming dwarfs, and tidal streamers of
stellar debris all suggest that the enhanced star formation rates (SFRs) in the two
main members may have been triggered by an earlier collision with another group
member.  It is unlikely that KPG103a and II Zwicky 23 have experienced a direct
collision with each other.  More likely they have been perturbed by a collision with a
dwarf companion or a weak interaction with each other, since there is evidence of disk
structure in both II Zwicky 23 and KPG 103a. 

  \section{II Zwicky 23 et al.: Family Traits}
  \label{wehner:properties}

  \subsection{II Zwicky 23}
  \label{wehner:IIzw23_properties}

II Zwicky 23 has $cz \sim 8320$ km s$^{-1}$ \cite{wehner:sargent70} and lies at a distance of 115
Mpc ($H_0 = 72$ km s$^{-1}$ Mpc$^{-1}$).  II Zwicky 23 qualifies as a luminous,
blue, compact galaxy, and its SFR is clearly enhanced.\cite{wehner:hg85} The 1.4 GHz radio
data \cite{wehner:condon02, wehner:ls04} suggest a SFR $\sim 4.65$ M$_{\odot}$ yr$^{-1}$ for M $>$ 5 
M$_{\odot}$ and including stars down to M $> 0.1$ M$_{\odot}$ brings the
SFR$_{1.4GHz}$ up to 25.8 M$_{\odot}$ yr$^{-1}$.  The far-infrared fluxes
\cite{wehner:kennicutt98, wehner:iras, wehner:ls04} yield a SFR $\sim 11.5$ M$_{\odot}$ yr$^{-1}$, and long
slit spectroscopy suggests SFR$_{H\alpha} = 4.5$ M$_{\odot}$ yr$^{-1}$\cite{wehner:ls04}.

II Zwicky 23 has an inclination of $i = 37$ degrees, and the northern side is moving away from us, while the southern side
is approaching us. \cite{wehner:ls04} In this case, the tidal debris tails emanating from the right of
II Zwicky 23 in Figure 1 are polar extensions, and the debris to the north is roughly
in the plane of the disk.

  \subsection{KPG103a}
  \label{wehner:kpg103a_properties}

KPG103a is the less well-studied companion of II Zwicky 23.  It lies 81 kpc to the
southwest and has a very similar velocity ($cz = 8313$ km s$^{-1}$) \cite{wehner:marzke96}.  It has
been classified an Sb, HII galaxy and is characterized by the presence of a weak,
two-armed spiral structure.\cite{wehner:marzke96} In optical images, there appear to be three
knots in the center of the galaxy, possibly the nucleus and a circumnuclear ring, or
multiple nuclei left over from an unrelaxed collision.  There also appear to be some
diffuse, low surface brightness extensions.  These properties, along with what appears
to be an enhanced SFR, suggest that KPG103a has recently undergone some type of tidal
interaction.

  \subsection{Tidal Debris, Dwarfs and Knots}
  \label{wehner:debris_properties}

A thorough analysis of the knots and regions of interest in II Zwicky 23 can be
found in \cite{wehner:ls04}.  We initially developed our own labeling
scheme, and our region labels are shown along with theirs in Figure 1.  
However, we have attempted to follow their numbering scheme whenever possible.  The
regions have been overlaid on our $B$-band image.  Regions 1 and 3 are tidal dwarf
galaxies (TDGs) and region 7 is most likely a giant HII region \cite{wehner:ls04}, rather
than an infalling galaxy, as originally proposed by \cite{wehner:keel88}.  
While not noted as being tidal dwarfs, regions 11 and 12,
along the polar extensions, do show signs of some new star formation in addition to
what appears to be material stripped from II Zwicky 23 \cite{wehner:ls04}.

Region N/K (hereafter, region N) has a distinct rotational pattern and a lower $[O/H]$ ratio than expected for
a tidally forming galaxy.   Therefore N is thought to be an independent group member, not formed as a result of any recent
collision\cite{wehner:ls04}.

\section{Observations and Reductions}
\label{wehner:obs_red}

We observed II Zwicky 23 and its companions using the MiniMosaic camera on the WIYN
3.5m telescope in Kitt Peak, Arizona.  Data were taken on two different runs.  On
November 17, 2000, we obtained broadband $B$ and $R$-band data.  We also observed II
Zwicky 23 on February 9, 2005 using the DensePak instrument at the Cassegrain focus of
the WIYN 3.5m telescope.  We obtained 2200s total, centered around the H$\alpha$ line. 

\section{Results}
\label{wehner:results}

  \subsection{Broadband Imaging}
  \label{wehner:bb_results}

In addition to those described in previous works (see Section 2), there are several new and
interesting features in our deep $B$-band image, shown in Figure 1.  The first is the
small, stellar arc, marked in Figure 1.  For an inclination of $37\deg$\cite{wehner:ls04},
the inner edge of this ripple lies approximately 18 kpc from the center of II Zwicky
23 and in the plane of the disk, and is therefore most likely a shell-like debris structure.  There is also a stellar extension that appears to
connect the center of II Zwicky 23 with the shell.  These stars could be the remains of
a disrupted companion galaxy, which plunged through II Zwicky 23 and created the
shell.  Considering the uncertainties in projection, it could extend out of the plane
of the disk, unconnected to the disk ripple.  It is interesting to note the presence
of stellar clumps further out, that lie along a line (as shown in Figure 1) that continue this stellar extension. 
These clumps lie at distances (uncorrected for projection) of 35 and 42 kpc from II
Zwicky 23. However, without further data, we cannot conclusively determine whether these clumps
are associated with this group.

%
%

\begin{figure}
\centering
\includegraphics[height=7cm]{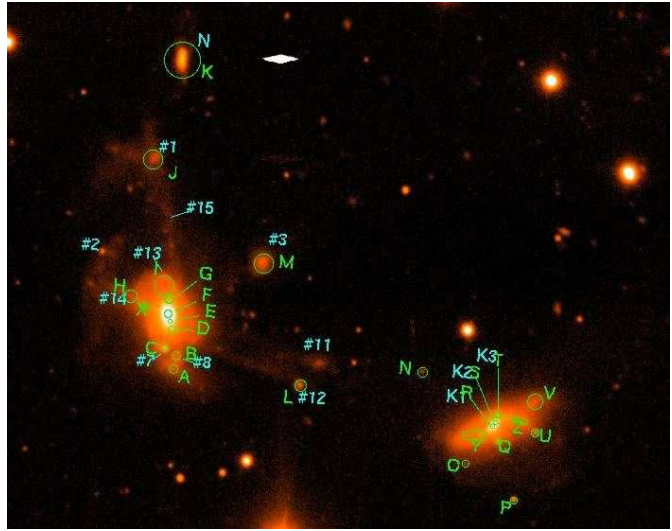}
\caption{$B$-band image from the WIYN 3.5m telescope in Kitt Peak, Arizona.  II
Zwicky 23 is on the left, and KPG103a is on the right.
Our region labels are alphabetical, while those region labels containing numbers
follow the notation in \cite{wehner:ls04}. }
\label{wehner_fig1}       
\end{figure}

KPG103a, also shown in Figure 1, exhibits many signatures of interaction.  In the
$B$-band image there appears to be a weak 2-armed spiral pattern, although it is
somewhat distorted. The northeastern arm curves briefly clockwise, before proceeding
to turn counter-clockwise.  While the other arm is more well-behaved, there is an
obvious diffuse stellar clump to its immediate upper right.  This clump could be the
remains of a merged galaxy or part of KPG103a's own disk that was pulled out by tidal
forces during an interaction.  This disk has clearly been disrupted by a collision
with another galaxy.

Also interesting are the two bright and roughly circular knots in the center of this
galaxy.  These knots may also contain the remains of a merged galaxy, and may even be
its nucleus.  KPG103a also has a larger, more diffuse low-surface brightness component
extending from the west/northwest.  This is most likely material pulled from KPG103a
during a tidal interaction.

Another feature of note is in the neighborhood of the dwarf galaxy, N.  In our deep
$B$-band image, there is a visible loop extending from region N to the north.  This
loop, while fainter in the shallower data, is visible in all our wavebands and is
therefore most likely a loop of stellar material.  One possible interpretation is that
region N is in the process of disrupting, leaving stars in its wake.  In this case,
the northern tip of the loop may be where N reached its peak distance from II Zwicky
23 and turned around in its orbit.  If we then use this peak, as its maximum orbital
distance from II Zwicky 23, and assume a simple orbit, then its last close encounter
with II Zwicky 23 was 1.4 Gyr ago - a time much too distant to have caused the current
central starburst, which is thought to be only $\sim 6$ Myr old. \cite{wehner:ls04}

  \subsection{DensePak}
  \label{dp_results}

The DensePak array contains 91 fibers, each of which spans 3 arcseconds.  The entire
array covers 30 $\times$ 40 arcseconds on the sky.  For each fiber with
detectable H$\alpha$, we fit a gaussian to the line to find the location of its peak.
From this, we calculated the velocity of the gas in each fiber.  We then assigned
the velocity range to a color table and plotted the velocity for each fiber, as shown in Figure 2.

%
%

\begin{figure}
\centering
\includegraphics[height=5.5cm]{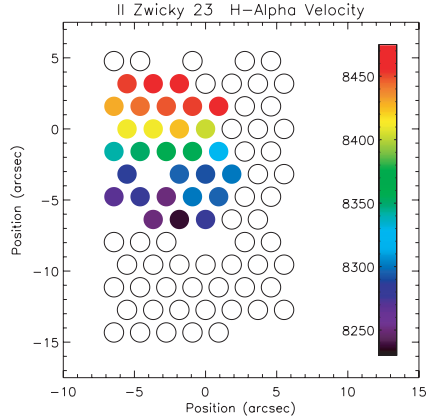}
\caption{DensePak data for II Zwicky 23 centered on the H$\alpha$ line.  The
velocity for each fiber is represented by its color. }
\label{wehner_fig2}       
\end{figure}

Despite its obvious signs of disturbance, we find a
fairly smoothly rotating disk, supporting II Zwicky 23's classification as a spiral
disk galaxy.  We measure a $\Delta v = 230$ km s$^{-1}$ across the disk, or a maximum
rotation speed of $v_{rot} \sim 117$ km s$^{-1}$.  It is possible that this number is an
overestimate, since it includes fibers such as that on region 7/C, which may represent 
the remains of an incoming, disrupting dwarf galaxy.\cite{wehner:keel88}

\section{Discussion}
\label{wehner:discussion}

II Zwicky 23 and its companions are a group in turmoil; there is now extensive
evidence that this system is being affected by numerous collisions.  The disruption of region
N, the ongoing starburst, and tidal arms of debris extending out of the disk all
suggest ongoing interactions between the members of this group.  In addition, the
presence of shells in II Zwicky 23, as well as the possible double nuclei in KPG103a
support the idea that collisions with former dwarf companions have recently
occurred.  However, despite all this, the ionized component of II Zwicky 23's disk
has maintained a smoothly rotating structure.  This system is
an excellent example of intra-group interactions driving galaxy evolution, and
merits future study.

%
%

\begin{thebibliography}{99.}

\bibitem{wehner:condon02} Condon, J. J., Cotton, W. D., \& Broderick, J. J. \ 2002, AJ, 124, 675
\bibitem{wehner:hg85} Hunter, D.A. \& Gallagher, J. S. \ 1985, AJ, 90, 1457
\bibitem{wehner:iras} Iras Point Source Catalog, \ 1996
\bibitem{wehner:keel88} Keel, W. C. \ 1988, A\&A, 202, 41
\bibitem{wehner:kennicutt98} Kennicutt, R. C. \ 1998, ApJ, 498, 541
\bibitem{wehner:landolt92} Landolt, A. \ 1992, AJ, 104, 340
\bibitem{wehner:ls04} L\'opez-S\'anchez, \'A. R., Esteban, C. \& Rodr\'iguez, M. \ 2004, A\&A, 428, 425
\bibitem{wehner:marzke96} Marzke, R. O., Huchra, J. P., \& Geller, M. J. \ 1996, AJ, 112, 1803
\bibitem{wehner:sargent70} Sargent, W. L. W. \ 1970, ApJ, 160, 405
\bibitem{wehner:sh05} Springel, V. \& Hernquist, L. \ 2005, ApJ, 622, L9

\end{thebibliography}
%
%
%



\printindex
\end{document}